\pdfoutput=1

\documentclass[a4paper]{article}

\usepackage{INTERSPEECH2022}

\usepackage{INTERSPEECH2022}
\usepackage{amsmath,graphicx, amsfonts}
\usepackage{multirow}
\usepackage{xcolor}
\usepackage{soul}
\usepackage{enumitem}
\usepackage[flushleft]{threeparttable}
\PassOptionsToPackage{hyphens}{url}\usepackage{hyperref}
\hypersetup{colorlinks=true}
\usepackage{changes}
\usepackage{todonotes}
\usepackage{adjustbox}

\usepackage{enumitem}
\setlist[itemize]{align=parleft,left=0pt..1em}

\usepackage{listings}
\usepackage{xcolor}

\definecolor{codegreen}{rgb}{0,0.6,0}
\definecolor{codegray}{rgb}{0.5,0.5,0.5}
\definecolor{codepurple}{rgb}{0.58,0,0.82}
\definecolor{backcolour}{rgb}{1.00,1.00,1.00}

\lstdefinestyle{mystyle}{
    backgroundcolor=\color{backcolour},   
    commentstyle=\color{codegreen},
    keywordstyle=\color{magenta},
    numberstyle=\tiny\color{codegray},
    stringstyle=\color{codepurple},
    basicstyle=\ttfamily\footnotesize,
    breakatwhitespace=false,         
    breaklines=true,                 
    captionpos=b,                    
    keepspaces=true,                 
    numbers=left,                    
    numbersep=5pt,                  
    showspaces=false,                
    showstringspaces=false,
    showtabs=false,                  
    tabsize=2
}

\title{4-bit Conformer with Native Quantization Aware Training for \\ Speech Recognition}
\name{Shaojin Ding, Phoenix Meadowlark, Yanzhang He, Lukasz Lew, Shivani Agrawal, Oleg Rybakov}
\address{Google LLC, USA}
\email{\{shaojinding,meadowlark,yanzhanghe,lew,shivaniagrawal,rybakov\}@google.com}

\begin{document}

\maketitle
\begin{abstract}

\noindent
Reducing the latency and model size has always been a significant research problem for live Automatic Speech Recognition (ASR) application scenarios. Along this direction, model quantization has become an increasingly popular approach to compress neural networks and reduce computation cost. Most of the existing practical ASR systems apply post-training 8-bit quantization. To achieve a higher compression rate without introducing additional performance regression, in this study, we propose to develop 4-bit ASR models with native quantization aware training, which leverages native integer operations to effectively optimize both training and inference. 
We conducted two experiments on state-of-the-art Conformer-based ASR models to evaluate our proposed quantization technique. First, we explored the impact of different precisions for both weight and activation quantization on the LibriSpeech dataset, and obtained a lossless 4-bit Conformer model with 5.8x size reduction compared to the float32 model. Following this, we for the first time investigated and revealed the viability of 4-bit quantization on a practical ASR system that is trained with large-scale datasets, and produced a lossless Conformer ASR model with mixed 4-bit and 8-bit weights that has 5x size reduction compared to the float32 model.

\end{abstract}

\section{Introduction}
% \footnote{Preprint. Submitted to INTERSPEECH}  % TODO remove
With the fast growth of voice search and speech-interactive features, automatic speech recognition (ASR)~\cite{wang2019overview, hannun2014deep, graves2012sequence, chorowski2015attention, dong2018speech} has become an essential component for user-interactive services and devices (e.g., search by voice functions in search engines and smartphones) over the years. Modern ASR applications are mostly developed based on an end-to-end model~\cite{li2020comparison,he2019streaming,CC18,KimHoriWatanabe17,JinyuLi2019,Zeyer2020}, which has been shown to achieve significant recognition performance improvements compared to conventional hybrid systems~\cite{Golan16} with a much smaller model size.

Improving latency and model size without compromising recognition quality has been an active research topic for years, as they benefit live ASR applications with both server-side and on-device models. Prior studies have explored the use of network pruning~\cite{takeda2017node, shangguan2019optimizing, gao2020rethinking, ding2021audio}, knowledge distillation~\cite{li2018compression}, and model quantization~\cite{han2015deep, alvarez2016efficient, he2019streaming, sainath2020streaming}. Among model quantization methods, post training quantization (PTQ) with int8~\cite{POSTQUANT} is popular and easy to use (e.g. by just setting a flag during model conversion in TFLite~\cite{POSTQUANT}). It is successfully applied in multiple applications~\cite{he2019streaming, sainath2020streaming}. One of the drawbacks of such technique is the potential performance degradation due to the loss of precision. Another limitation of PTQ is the lack of control over model quantization, e.g. it does not support int4 quantization or customized quantization of selected set of layers. That is why in this work we are focused on quantization aware training (QAT).

QAT has several flavors: ``fake"~\cite{Benoit} and native, discussed in~\cite{abdolrashidi2021pareto} also called Accurate Quantized Training. It is called accurate because it uses native integer operations for executing quantized operations (e.g. matrix multiplications) and does not have any difference between accuracy during training and inference. Whereas ``fake quantization" can have a numerical difference between training (with float operations) and inference (with integer operations) modes if float operation does not fit into 23 bits of mantissa during training. 

In this work we use native QAT for ASR model, because it follows the approach of ``what you train is what you serve". It can use native integer operations during training (if hardware supports it or else it will use int emulated in float32, A.k.a. ``fake quantization"). With native integer operations, there is no numerical difference between forward propagation of training and inference. 
We conduct extensive experiments on LibriSpeech and large-scale Voice Search datasets with a state-of-the-art Conformer ASR model~\cite{gulati2020conformer} to evaluate this quantization approach, and systematically analyze how different quantization precisions affect the compute cost-accuracy tradeoff with the main focus on int4\footnote{We interchangeably use 4-bit and int4 (8-bit and int8) hereafter.} quantization. Our main contributions are outlined as below:

\begin{itemize}[leftmargin=*]
    \item We leverage native QAT approach for ASR model, instead of running ``fake quantization" that is mostly used in previous studies~\cite{fasoli20214, bie2019simplified, kim2021integer}. It allows us to run the model on both cloud (e.g. on TPU) and mobile applications, whereas ``fake" QAT is used mostly for mobile applications and needs special conversion done by  TFlite~\cite{POSTQUANT}.
    \item We minimize the number of operations used for quantization; so that training time with native QAT is increased only by 7\% (on TPU that supports float operations only) in comparison to training of a float model. If native QAT is executed on hardware that supports integer operations, then we expect more speed up in training time.
    \item We demonstrate that \textit{Large} Conformer ASR model on Librispeech data has minimal or no accuracy loss with int4 weights and float32 activations quantization, or int4 weights and int8 activations quantization (as expected it also works well with int8 weights and activation quantization). For the first time we evaluate native 4-bit QAT on a real production model trained on large scale data. Despite of no accuracy regression on public Librispeech data, we observe that native QAT introduces regression on the model trained with large data sets. More importantly, we investigate several strategies to mitigate the regression and affect the cost-accuracy trade off, bringing new insights to both model quantization and ASR modeling studies.
\end{itemize}

\textbf{Relation to prior work.} A number of previous studies have also explored the idea of ASR model quantization~\cite{fasoli20214, bie2019simplified, kim2021integer}. Our work differs from them in several aspects. First and foremost, most previous studies conduct experiments only on relatively small datasets (e.g., LibriSpeech, Switchboard, and CallHome). The models trained on these datasets are usually heavily over-parameterized, which makes it easy to obtain a loss-less 8/4-bit quantized model. By contrast, we for the first time examine 4-bit ASR quantization on a model trained on a large scale dataset ($\sim$400k hours) and showed the behavior difference. Second, all three studies use ``fake quantization" that can have a numerical difference between training and inference. However, in this study, we consider native QAT, which has several advantages: 1) the same model representation can be used for both mobile and cloud model training and inference; 2) if hardware supports integer operations then we can expect model training speed up. Lastly,~\cite{bie2019simplified, kim2021integer} only explore 8- or 6-bit quantizations. Although~\cite{fasoli20214} also explored 4-bit quantization that is similar to ours, there are several differences: 1) we focus on a recent state-of-the-art Conformer-based model, but~\cite{fasoli20214} uses LSTM ASR; 2) we train model with native QAT from scratch, whereas in~\cite{fasoli20214} they first train float model and then fine-tune it with QAT, which complicates training pipeline and increases total training time; 3) to minimize the impact of native QAT on total training time we reduced the number of operations in quantization function and do not have clamp with learnable boundaries as in~\cite{fasoli20214}.

\section{Method}

\begin{figure}[t]
    \centering
    \includegraphics[width=\columnwidth]{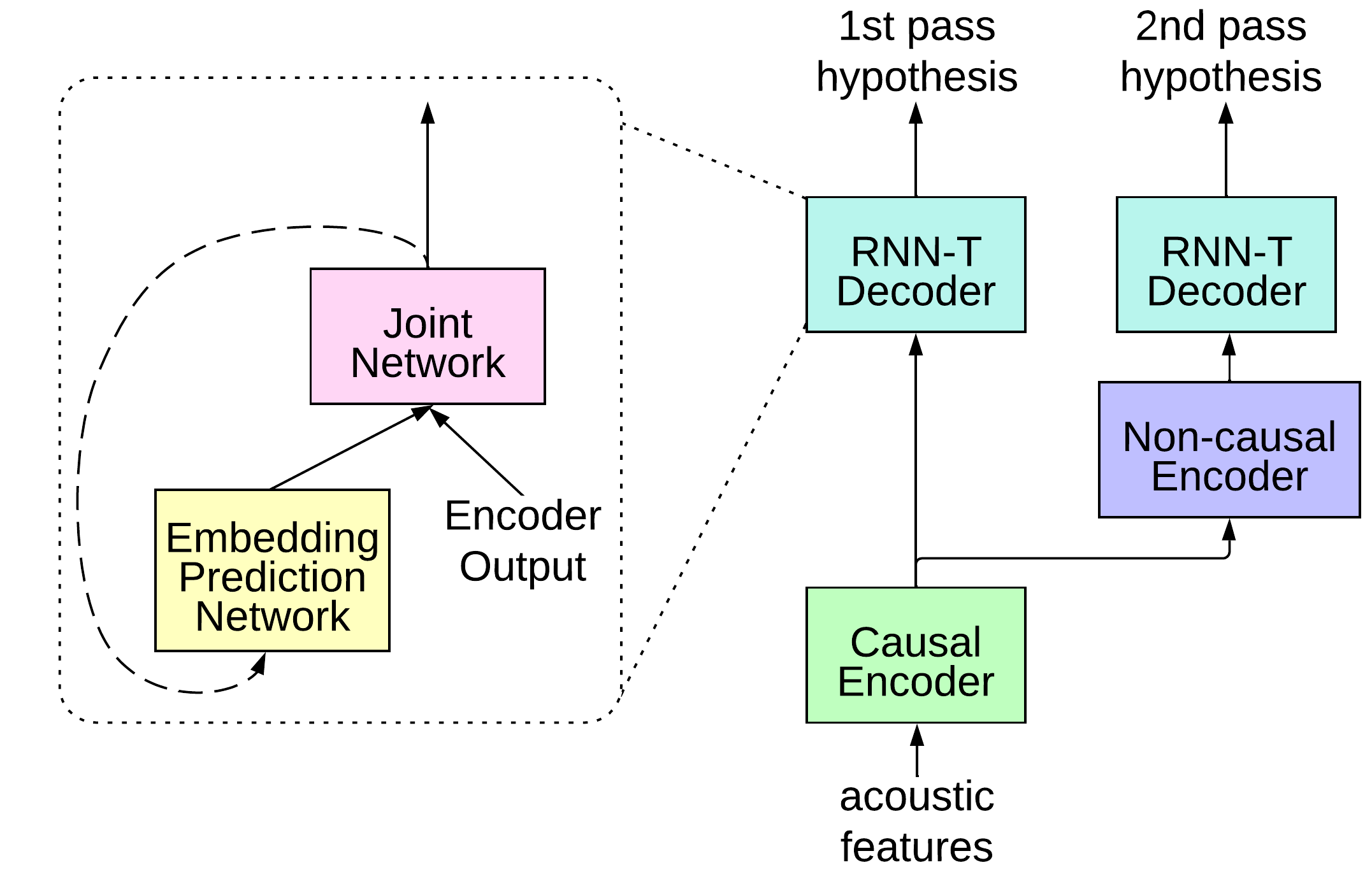}
    \caption{The architecture for the multi-pass ASR model.}
    \label{fig:cascade}
    \vspace{-0.1in}
\end{figure}

\subsection{ASR model architectures}
\label{sec:backbone}
We use the state-of-the-art Conformer Transducer~\cite{gulati2020conformer} backbones in this study. We considered slightly different architectures for the LibriSpeech experiments and the production experiments based on a real-world on-device ASR system to guarantee fair comparisons to the corresponding baselines. For LibriSpeech, we follow the architecture proposed in~\cite{gulati2020conformer}, which has a single encoder with different layers for \textit{Small} (10M parameters) and \textit{Large} (118M) models. The decoder is a standard RNN-Transducer decoder with 1-layer LSTM.

Alternatively, for model in practical applications, we follow the multi-pass architecture proposed in~\cite{sainath2021cascadedlm} and shown on Figure~\ref{fig:cascade}. It has a first pass causal encoder (47M parameters), followed by a second pass non-causal encoder (60M). This multi-pass model
% \Ryan{never introduced before} 
unifies the streaming and non-streaming ASRs, where the causal encoder uses only left context and produces partial results with minimal latency, and the non-causal can provide more accurate hypothesis by using both left and right context. Unlike the original architecture, we use a separated decoder (4.4M) for each encoder. The decoder has an embedding prediction network~\cite{Rami21} and a 1-layer fully-connected joint network. The embedding prediction network further reduces the decoder model size without accuracy degradation.

\begin{figure}[t]
  \centering
\begin{lstlisting}[language=python]
def quantize(input, axis):
  max_val = tf.math.reduce_max(tf.math.abs(input), axis=axis, keepdims=True)
  scale = tf.divide(max_val, 127.0)
  scale = tf.stop_gradient(scale)
  scaled_input = tf.math.divide_no_nan(input, scale)
  return tf.cast(
      tf.math.floor(scaled_input + 0.5), tf.int8)
\end{lstlisting}
  \caption{int8 native quantization in TensorFlow.}
  \label{fig:quant}
  \vspace{-6mm}
\end{figure}

\subsection{Native quantization aware training}
\label{sec:native_qat}
Standard approach for QAT in TensorFlow(TF)~\cite{TF} is based on ``fake" QAT and relies on  tf.quantization.fake\_quant\_* operations. In this case, researchers have to use these operations during training and server-side inference. For on-device models, they use TFlite for converting fake quantization operations to integer operations
% \Ryan{incomplete?} 
and then run output model with TFLite on mobile phone. Even though this approach provides end-to-end user experience, it also has some disadvantages: 1) It requires additional conversion step of fake quantization operations to integer. 2) Existing API tf.quantization.fake\_quant\_with\_min\_max\_vars\_per\_channel supports per-channel min and max values estimation over the last dimension only. However, there are use cases when channel dimension is not the last one. In this case we have to permute dimensions of the input tensor to make channel dimension the last one, then we ``fake" quantize the tensor with tf.quantization.fake\_quant\_with\_min\_max\_vars\_per\_channel and after that permute dimensions back to the original order of the input tensor. These additional permutation operations can increase training time in comparison to the non-quantized model training time.
To address all above we propose to use native QAT and with native tf operations, also discussed in \cite{abdolrashidi2021pareto}; called Accurate Quantized Training. It allows us to follow an approach called ``what you train is what you serve". As a result we can use the same TF model for training and inference on both mobile and cloud TPU applications with opportunity to speed up model training if hardware supports integer operations (e.g. matrix multiplications).

One of the methods of 4-bit ASR model quantization is based on QAT fine-tuning \cite{fasoli20214}. It has several steps: train float model, then use it for QAT fine tuning. This approach always increases total training time (it includes float model training and QAT fine tuning). By contrast, we train our model from scratch in quantization aware mode and show that training time is increased by only 7\% (on hardware with float operations). We achieve it by using native QAT based on dynamic quantization shown in Figure~\ref{fig:quant} (for int8 use case). It estimates max values over axis which will be quantized (it supports channel-wise quantization). Then it computes \textit{scale} by dividing max values with 127 for int8 (or by 7 for int4). After that, the input tensor is quantized by dividing with \textit{scale} and casting it to int. De-quantization will be done by multiplying a tensor with \textit{scale}. To reduce overall computations we do not use ``zero point" as in \cite{Benoit} and assume that input tensor values distribution is symmetrical. Even though it is a strong assumption, in Section~\ref{sec:exp_libri}, we empirically show that model weights can be quantized by native 4-bit QAT without accuracy degradation.

\section{Experimental setups}

\subsection{Datasets}

We conducted experiments on both LibriSpeech and an internal large-scale datasets. LibriSpeech training set contains 960 hours of speech, where 460 hours of them are “clean” speech and the other 500 hours are “noisy” speech. The testing set also consists of a “clean” and a “noisy” subset. 

When running experiments with the large-scale datasets, we train the models with a training set~\cite{narayanan2019recognizing, sainath2020streaming} consisting of $\sim$400k hours English audio-text pairs from multiple domains, such as YouTube and anonymized voice search traffic. YouTube were transcripbed in a semi-supervised fashion~\cite{liao2013large}. All other domains are anonymized and hand-transcribed. During evaluations, we use the Voice Search (VS) test set that contains around 12k voice search utterances, each having an average length of 5.5 seconds. Our data handling abides by \textit{Google AI Principles}~\cite{googleaiprinciples}.

\subsection{Conformer model with Librispeech data}

The LibriSpeech conformer model~\cite{gulati2020conformer} uses a frontend of 80-dimensional log Mel-filterbank energies, extracted from 25ms window and 10ms shift. The \textit{Small}(S) and \textit{Large}(L) variants have 16,  17 layers, with a dimensionality of 144, 512, respectively. The \textit{Small} model has 4 attention heads in self-attention layers, while the \textit{Large} model has 8 heads. The kernel size of the depthwise convolutions is set to 32. The LSTM layer in decoder has 640 units. 

\subsection{Conformer model with large-scale data}

The model with large-scale data uses a 128-dimensional log Mel-filterbank enegies as the frontend feature, in which the 4 contiguous frames are stacked, and the stacked sequence is sub-sampled by a factor of 3. We use causal convolution for all layers with a kernel size of 15. The causal conformer encoder has 7 conformer layers (first 3 layers have no self-attention), which has 23-frame left context per layer and no right context to strictly prevent the model from using future inputs. The non-causal encoder 6 conformer layers, with additional 30-frame right context across 6 layers that processes 900ms speech from the future. All the self-attention layers have 8 heads. Each separate RNN-T decoder is comprised of an 320-dimensional embedding prediction network and a 384-dimensional fully-connected joint network.

Evaluations of production models are running on an on-device inference pipeline, where we convert the TensorFlow graphs to TFLite format, with the corresponding quantization approach. Additionally, we did not use any language model in our experiments, as this is orthogonal to the end-to-end model performance. When applying native QAT to the backbone models, we only quantize the encoders since the decoders are relatively small. We do not quantize convolutional kernels for the same reason.

\section{Results}

We conduct two sets of experiments to evaluate the proposed native QAT on ASR models. First, we examine different quantization precision and compared with five baselines on LibriSpeech dataset, demonstrating the validity and the effectiveness of native QAT in conformer based ASR models. Second, we explore applying native QAT to a pratical ASR models that is trained on large-scale data and the corresponding optimizations.

\subsection{Experiments on LibriSpeech}
\label{sec:exp_libri}

\begin{table}[t]
\begin{center}
\caption{Results of our proposed int8/4 QAT on Conformer Large(L) and Small(S) models with the baseline approach~\cite{kim2021integer} on LibriSpeech test-clean and test-other subsets. Please see Section~\ref{sec:exp_libri} for the meanings of the method abbreviations.}
\label{table:librispeech}
\resizebox{\columnwidth}{!}{
\begin{tabular}{l|cc|cc|cc}
\hline
\multicolumn{4}{c}{Conformer (L)} \\
 \hline
 \textbf{Method} & \textbf{test-clean} & \textbf{test-other} & \textbf{Model size (MB)} \\
 \hline
 Float & 2.0 & 4.4 & 475 \\
 \hline
 I8W & 2.0 & 4.5 & 139 \\
 I4W & 2.0 & 4.4 & 82 \\
 I8WA & 2.0 & 4.5 & 139 \\
 I4WI8A & 2.1 & 4.4 & 82 \\
 I4WA & 3.1 & 8.2 & 82 \\
\hline
 FakeI4W & 2.0 & 4.6 & 82 \\
\hline
\hline
\multicolumn{4}{c}{Conformer (S)} \\
 \hline
 Float & 2.5 & 6.1 & 42\\
 \hline
 I8W & 2.5 & 6.0 & 17\\
 I4W & 2.7 & 6.3 & 13\\
 I8WA & 2.5 & 6.0 & 17\\
 I4WI8A & 2.8 & 6.6 & 13\\
 I4WA & 5.0 & 12.1 & 13 \\
\hline
 FakeI4W & 2.9 & 6.9 & 13\\
  \hline
   \hline
\multicolumn{4}{c}{Baselines} \\
\hline
 \cite{prasad2020quantization}I8WA & 6.9 & N/A & 8 \\
 \cite{nguyen2020quantization}I8W & 8.7 & 22.3 & 60 \\
 \cite{nguyen2020quantization}I6W & 8.9 & 22.8 & 45 \\
 \cite{kim2021integer}I8W & 2.7 & 6.9 & 123 \\
 \cite{kim2021integer}I6W8A & 3.6 & 8.2 & 92  \\
   \hline
\end{tabular}}
\vspace{-10pt}
\end{center}
\end{table}

We experiment with conformer \textit{Large} and \textit{Small} models to examine the behaviors of QAT with different model sizes. In terms of QAT, we consider 7 quantization configurations to understand the impacts of 8/4-bit quantization on weights only or both weights and activations:

\begin{itemize}
    \item \textit{Float}: float32 weight, float32 activation (baseline)
    \item \textit{I8W}: int8 weight, float32 activation
    \item \textit{I4W}: int4 weight, float32 activation
    \item \textit{I8WA}: int8 weight, int8 activation
    \item \textit{I4WI8A}: int4 weight, int8 activation
    \item \textit{I4WA}: int4 weight, int4 activation
    \item \textit{FakeI4W}: fake int4 weighs, float32 activation
\end{itemize}

\noindent
Quantizing the weights alone can reduce the model size, but we still need to convert the weights to float during inference. By contrast, quantizing the activations will benefit in both model size and run-time efficiency, with integer matrix multiplication.

We train models with different combinations of weight and activation quantization using native QAT and ``fake" QAT approaches, and reported their WERs in Table~\ref{table:librispeech}.  Here, we first analyze the impact of quantizing only model weights to 8/4-bit. In 8-bit cases (\textit{I8W}), we see no regression in either \textit{Large} or \textit{Small} model. However, in 4-bit cases (\textit{I4W}), although the \textit{Large} model can still retain the float performance, the \textit{Small} model has introduced 0.2 WER increase compared to the baseline due to the limited capacity of the \textit{Small} model. Additionally, we also investigate the possibility of quantizing both weights and activations. With \textit{I8WA}, the \textit{Large} and \textit{Small} models can still preserve the float performance. As we quantized the model more aggressively with \textit{I4WI8A} and \textit{I4WA}, we see more serious performance loss. To summarize, the most light-weight configurations without WER loss of \textit{Large} model is to have int4 weights and int8 activations. For \textit{Small} model, the most light-weight configuration without WER loss is to use int8 for both weights and activations. Compared against the baseline models, our most light-weight \textit{I4WI8A} Conformer (S) model has already significantly outperformed all the baselines while having more aggressive quantization scheme, demonstrating our state-of-the-art performance.

``Fake" QAT with 4-bit weights has the same accuracy as the float model on \textit{test-clean} and little accuracy reduction on \textit{test-other} as shown on Table~\ref{table:librispeech}. We benchmarked ``fake" QAT \textit{Fake4W} and native QAT \textit{I4W} with~\cite{shen2019lingvo} on TPU and observed that training with native QAT is 7\% slower than float model. We explain it by additional operations shown in Figure ~\ref{fig:quant} and by the fact that we used hardware with float operations support only (we expect speedup on hardware with integer operations support). We also observed that training with \textit{Fake4W} is 6\% slower than native QAT.

\subsection{Exploring the limit of 4-bit quantization on large-scale data}
\label{sec:exp_prod}

\begin{table}[t]
\begin{center}
\caption{Results of applying int8/4 QAT to production ASR model. We use the actual TensorFlow Lite file size of the models with corresponding quantization approaches for model size, measure in megabyte (MB). PTQ refers to post-training quantization. From model \textit{E4} to \textit{E13}, we apply in4 QAT to the listed layers and int8 PTQ for the remaining layers.}
\label{table:prod_model}
\resizebox{\columnwidth}{!}{
\begin{tabular}{c|l|cc|c}
\hline
\multirow{2}{*}{Exp} & \multicolumn{1}{c|}{\multirow{2}{*}{Model}} & \multicolumn{2}{c|}{Voice Search WER} & Model size  \\
 \cline{3-4} & &  1st pass & 2nd pass & (MB) \\
 \hline
 B0 & float32 model & 8.0 & 5.7 & 460 \\
  \hline
 E0 & int8 PTQ & 7.9 & 5.8 & 115 \\
 E1 & int8 QAT & 7.9 & 5.8 & 115\\
  \hline
 E2 & int4 PTQ & 17.7 & 16.0 & 61 \\
 E3 & int4 QAT & 8.5 & 6.2 & 61 \\
 E4 & int4 QAT causal & 8.5 & 5.8 & 91 \\
 E5 & int4 QAT non-causal & 7.9 & 6.2 & 84 \\
 \hline
 E6 & int4 except first\& last & 8.5 & 6.1 & 78 \\
 E7 & int4 except self-atten & 8.1 & 5.9 & 94 \\
 \hline
 \multicolumn{5}{c}{Quantizing different number of layers in both encoders} \\
 \hline
 E8 & int4 first layer & 7.9 & 5.8 & 108 \\
 E9 & int4 first 2 layer & 7.9 & 5.8 & 101 \\
 E10 & int4 first 3 layer & 7.9 & 5.9 & 94 \\
 E11 & int4 first 4 layer & 8.0 & 6.1 & 84 \\
 E12 & int4 first 5 layer & 8.1 & 6.2 & 75 \\
 E13 & int4 first 6 layer & 8.5 & 6.2 & 66 \\
 \hline

\hline
%  \cline{3-6} & VS &  \\
\end{tabular}}
\vspace{-20pt}
\end{center}
\end{table}

Although we can easily obtain int4 quantized model that has no performance regression in LibriSpeech, we still see performance drops when applying native int4 QAT to the model trained with large-scale data, even only with weight quantization. Results are shown in Table~\ref{table:prod_model}. Comparing to float32 models (\textit{B0}), the int8 models (\textit{E0} and \textit{E1}) have similar results. Unless otherwise indicated, \textit{we compare the following 4-bit models with E0}, since int8 PTQ has been widely used in real-world applications. For int4 models, with PTQ (\textit{E2}), the model cannot produce a reasonable performance anymore. By contrast, native QAT (\textit{E3}) has additional 0.6/0.4 degradations compared to int8 models, which again confirms the effectiveness of native QAT. The reason of the different conclusions between LibriSpeech and production models could be that LibriSpeech models are highly over-parameterized (LibriSpeech training set: 960 hours; large-scale training set: 400k hours), and therefore, the models still have enough capacities with 4-bit weights. By contrast, the production model is trained with an extremely large amount of data, so the model is not overfitting anymore. This can also be verified through the int8 results -- when applying native QAT to int8 model, we did not see any performance improvement over PTQ.

Additionally, with the multi-pass encoder architecture, we explore only quantizing either the first pass causal encoder or the second pass non-causal encoder. \textit{We keep other layers to be int8 PTQ hereafter.} As shown in Table~\ref{table:prod_model} (\textit{E4} and \textit{E5}), this keeps the performance of the non-quantized pass unchanged, which can be an intermediate solution for the scenarios where either the first or second pass is more important (e.g., we are more tolerant to regressions on the first pass if only using it for partial hypothesis). 

We further investigate the possibility of having no performance loss in either pass with three strategies. First, we keep the first and the last layer to be int8, as the they tend to be most sensitive to quantization~\cite{abdolrashidi2021pareto, choi2018pact}. However, as shown in Table~\ref{table:prod_model}, this observation does not hold for our model (\textit{E6}), and the performance remains to be the same as quantizing the entire causal encoder to int4. Second, we keep the self-attention layers to be int8 since they are the most essential layers in conformer. With this, we obtain a model (\textit{E7}) with 0.2/0.1 WER regression for the two passes compared to the int8 model. Lastly, we experiment with quantizing different number of layers from bottom to the top in both passes (\textit{E8} to \textit{E13}). When quantizing the first 3 layers of both encoders (\textit{E10}), we only see 0.1 performance loss on the 2nd pass. When quantizing additional layers, the model starts to have more serious degradations on both passes. These results indicate that lower layers are more robust to quantization, as the upper layers can mitigate the performance loss. It is also worthwhile to experiment with quantizing different number of layers in the two decoders, which we will continue investigating in future work. 

Consequently, we find that model \textit{E10} has the best tradeoffs, mostly retaining the performance of int8 or float32 models. In terms of model size, our int4 model has 81\% of int8 model size and only 20\% of the original float32 model size, which can significantly reduce memory, latency, and power consumption for on-device and server applications.

\section{Conclusions}

In this paper, we proposed a novel approach based on native QAT to establish state-of-the-art 4-bit quantized ASR model. Through an experiment on LibriSpeech, we analyzed the impact of different quantization configurations for both weights and activations, validating the effectiveness of QAT in building 4-bit quantized ASR models. Following this, we for the first time applied it to a production ASR with large-scale data. Observing a performance regression on production ASR models, we proposed three strategies to alleviate the regression and thoroughly discussed the cost-accuracy tradeoffs.

\bibliographystyle{IEEEtran}

\bibliography{main}

\end{document}